\documentclass[prd,preprint,showpacs]{revtex4}
\usepackage{amsmath}
\usepackage{latexsym}
\usepackage{amsfonts}
\usepackage{graphicx}
\begin{document}

\title{Growth factor parametrization and modified gravity}
\author{Yungui Gong}
\email{gongyg@cqupt.edu.cn}

\affiliation{College of Mathematics and Physics, Chongqing
University of Posts and Telecommunications, Chongqing 400065, China}

\begin{abstract}
The growth rate of matter perturbation and the expansion rate of the
Universe can be used to distinguish modified gravity and dark energy
models in explaining the cosmic acceleration. The growth rate is
parametrized by the growth index $\gamma$. We discuss the dependence
of $\gamma$ on the matter energy density $\Omega$ and its current
value $\Omega_0$ for a more accurate approximation of the growth
factor. The observational data, including the data of the growth
rate, are used to fit different models. The data strongly disfavor
the Dvali-Gabadadze-Porrati model. For the dark energy model with a
constant equation of state, we find that $\Omega_0=0.27\pm 0.02$ and
$w=-0.97\pm 0.09$. For the $\Lambda$CDM model, we find that
$\gamma=0.64^{+0.17}_{-0.15}$. For the Dvali-Gabadadze-Porrati
model, we find that $\gamma=0.55^{+0.14}_{-0.13}$.
\end{abstract}

\preprint{arXiv:0808.1316}
\maketitle
\section{Introduction}

The discovery of late time cosmic acceleration \cite{acc1}
challenges our understanding of the standard models of gravity and
particle physics. Within the framework of Friedmann-Robertson-Walker
cosmology, an exotic energy ingredient with negative pressure,
dubbed dark energy, is invoked to explain the observed accelerated
expansion of the Universe. One simple candidate of dark energy which
is consistent with current observations is the cosmological
constant. Because of the many orders of magnitude discrepancy
between the theoretical predication and the observation of vacuum
energy, other dynamical dark energy models were proposed \cite{rev}.
By choosing a suitable equation of state $w=p/\rho$ for dark energy,
we can recover the observed expansion rate $H(z)$ and the luminosity
distance redshift relation $d_L(z)$. Current observations are unable
to distinguish many different dark energy models which give the same
$d_L(z)$, and the nature of dark energy is still a mystery. Many
parametric and nonparametric model-independent methods were proposed
to study the property of dark energy
\cite{astier01,huterer,par2,par3,lind,alam,jbp,par4,par1,
par5,gong04,gong05,gong06,gong08,sahni}.

Bear in mind that the only observable effect of dark energy is
through gravitational interaction; it is also possible that the
accelerated expansion is caused by modification of gravitation. One
example of the alternative approach is provided by the
Dvali-Gabadadze-Porrati (DGP) brane-world model \cite{dgp}, in which
gravity appears four dimensional at short distances but is modified
at large distances. The question one faces is how to distinguish
such an alternative approach from the one involving dark energy. One
may answer the question by seeking a more accurate observation of
the cosmic expansion history, but this will not break the
degeneracies between different approaches of explaining the cosmic
acceleration. Recently, it was proposed to use the growth rate of
large scale in the Universe to distinguish the effect of modified
gravity from that of dark energy. While different models give the
same late time accelerated expansion, the growth of matter
perturbation they produce differ \cite{jetp}. To linear order of
perturbation, at large scales the matter density perturbation
$\delta=\delta\rho/\rho$ satisfies the following equation:
\begin{equation}
\label{denpert}
\ddot{\delta}+2H\dot{\delta}-4\pi G_{eff}\,\rho\delta=0,
\end{equation}
where $\rho$ is the matter energy density and $G_{eff}$ denotes the
effect of modified gravity. For example,
$G_{eff}/G=(4+2\omega)/(3+2\omega)$ for the Brans-Dicke theory
\cite{boiss} and $G_{eff}/G=2(1+2\Omega^2)/3(1+\Omega^2)$ for the
DGP model \cite{lue}, the dimensionless matter energy density
$\Omega=8\pi G\rho/(3H^2)$. In terms of the growth factor
$f=d\ln\delta/d\ln a$, the matter density perturbation Eq.
(\ref{denpert}) becomes
\begin{equation}
\label{grwthfeq1}
f'+f^2+\left(\frac{\dot{H}}{H^2}+2\right)f=\frac{3}{2}\frac{G_{eff}}{G}\Omega,
\end{equation}
where $f'=df/d\ln a$. In general, there is no analytical solution to
Eq. (\ref{grwthfeq1}), and we need to solve Eq. (\ref{grwthfeq1})
numerically; it is very interesting that the solution of the
equation can be approximated as $f=\Omega^\gamma$
\cite{peebles,fry,lightman,wang} and the growth index $\gamma$ can
be obtained for some general models. The approximation was first
proposed by Peebles for the matter dominated universe as
$f(z=0)=\Omega^{0.6}_0$ \cite{peebles}; then a more accurate
approximation, $f(z=0)=\Omega^{4/7}_0$, for the same model was
derived in \cite{fry,lightman}. For the Universe with a cosmological
constant, the approximation
$f(z=0)=\Omega^{0.6}_0+\Omega_{\Lambda0}(1+\Omega_0/2)/70$ can be
made \cite{lahav}. For a dynamical dark energy model with slowly
varying $w$ and zero curvature, the approximation
$f(z)=\Omega(z)^\gamma$ was given in \cite{wang}. For the DGP model,
$\gamma=11/16$ \cite{linder07}. Therefore, instead of looking for
the growth factor by numerically solving Eq. (\ref{grwthfeq1}), the
growth index $\gamma$ may be used as the signature of modified
gravity and dark energy models. It was found that
$\gamma=0.55+0.05[1+w(z=1)]$ with $w>-1$ and
$\gamma=0.55+0.02[1+w(z=1)]$ with $w<-1$ for the dynamical dark
energy model in flat space \cite{linder05,linder07}. Recently, the
use of the growth rate of matter perturbation in addition to the
expansion history of the Universe to differentiate dark energy
models and modified gravity attracted much attention
\cite{huterer07,sereno,knox,ishak,yun,polarski,sapone,balles,gannouji,berts,laszlo,kunz,kiakotou,porto,wei,ness}.

The dependence of $\gamma$ on the equation of state $w$ has received
much attention in the literature; we discuss the dependence of
$\gamma$ on $\Omega$ and $\Omega_0$ for a more accurate
approximation in this paper. We discuss a more accurate
approximation for the dark energy model with constant $w$ in Sec.
II. Then we discuss the DGP model in Sec. III. In Sec. IV, we apply
the Union compilation of type Ia supernovae (SNe) data \cite{union},
the baryon acoustic oscillation (BAO) measurement from the Sloan
Digital Sky Survey (SDSS) \cite{sdss6}, the shift parameter measured
from the Wilkinson Microwave Anisotropy Probe 5 yr data (WMAP5)
\cite{wmap5}, the Hubble parameter data $H(z)$ \cite{hz1,hz2}, and
the growth factor data $f(z)$ \cite{porto,ness,guzzo} to constrain
the models. We also use the growth factor data alone to find out the
constraint on the growth index $\gamma$. We conclude the paper in
Sec. V.

\section{$\Lambda$CDM model}
\begin{figure}[htp]
\centering
\includegraphics[width=12cm]{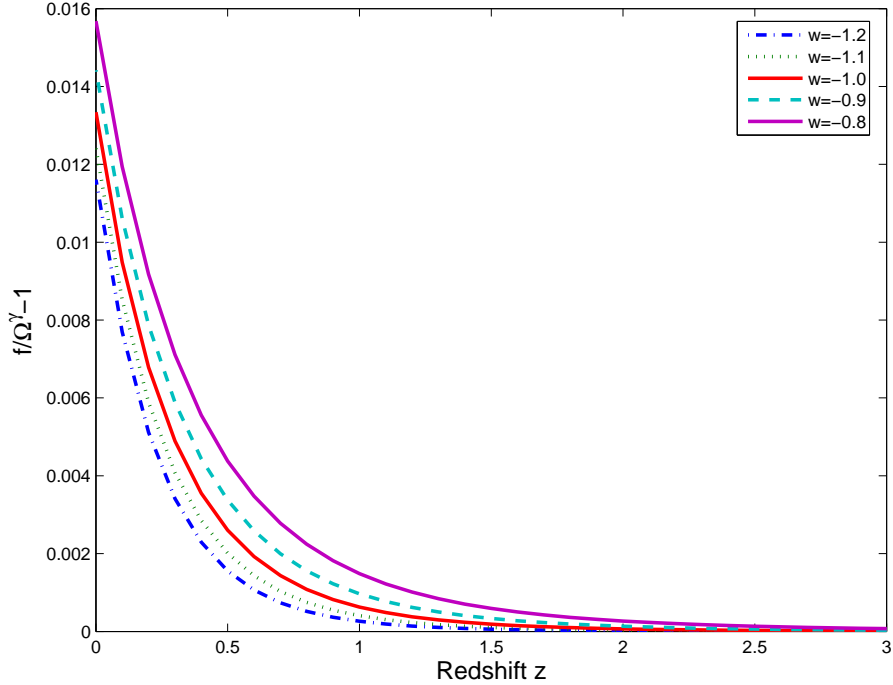}
\caption{The relative difference between the growth factor $f$ and $\Omega^{\gamma_\infty}$ with $\gamma_\infty=3(1-w)(5-6w)$
for the dark energy model with constant $w$.}
\label{wcdmferr}
\end{figure}

We first review the derivation of $\gamma$ given in \cite{wang}. For
the flat dark energy model with constant equation of state $w$, we
have
\begin{equation}
\label{wcdmhdot}
\frac{\dot H}{H^2}=-\frac{3}{2}[1+w(1-\Omega)].
\end{equation}
The energy conservation equation tells us that
\begin{equation}
\label{wcdmwmder}
\Omega'=3w\Omega(1-\Omega).
\end{equation}
Substituting Eqs. (\ref{wcdmhdot}) and (\ref{wcdmwmder}) into Eq. (\ref{grwthfeq1}), we get
\begin{equation}
\label{wcdmfeq}
3w\Omega(1-\Omega)\frac{df}{d\Omega}+f^2+\left[\frac{1}{2}-\frac{3}{2}w(1-\Omega)\right]f=\frac{3}{2}\Omega.
\end{equation}
Plugging $f=\Omega^\gamma$ into Eq. (\ref{wcdmfeq}), we get
\begin{equation}
\label{wcdmfeq1}
3w\Omega(1-\Omega)\ln\Omega\frac{d\gamma}{d\Omega}-3w\Omega(\gamma-1/2)+\Omega^\gamma-\frac{3}{2}\Omega^{1-\gamma}
+3w\gamma-\frac{3}{2}w+\frac{1}{2}=0.
\end{equation}
Expanding Eq. (\ref{wcdmfeq1}) around $\Omega=1$, to the first order
of $(1-\Omega)$, we get \cite{wang}
\begin{equation}
\label{wcdmr}
\gamma=\frac{3(1-w)}{5-6w}+\frac{3}{125}\frac{(1-w)(1-3w/2)}{(1-6w/5)^2}(1-\Omega).
\end{equation}
In order to see how well the approximation $\Omega^\gamma$ fits the
growth factor $f$, we need to solve Eq. (\ref{wcdmfeq1}) numerically
with the expression of $\Omega$. The dimensionless matter density is
\begin{equation}
\label{wcdmwm}
\Omega=\frac{\Omega_0 }{\Omega_0+(1-\Omega_0)(1+z)^{3w}}.
\end{equation}
Since $\gamma$ does not change very much, we first use
$\gamma=\gamma_\infty=3(1-w)/(5-6w)$ to approximate the growth
factor. For convenience, we choose $\Omega_0=0.27$, and the result
is shown in Fig. \ref{wcdmferr}.
From Fig. \ref{wcdmferr}, we see that $\Omega^{\gamma_\infty}$
approximates $f$ better than 2\%.

\begin{figure}[htp]
\centering
\includegraphics[width=12cm]{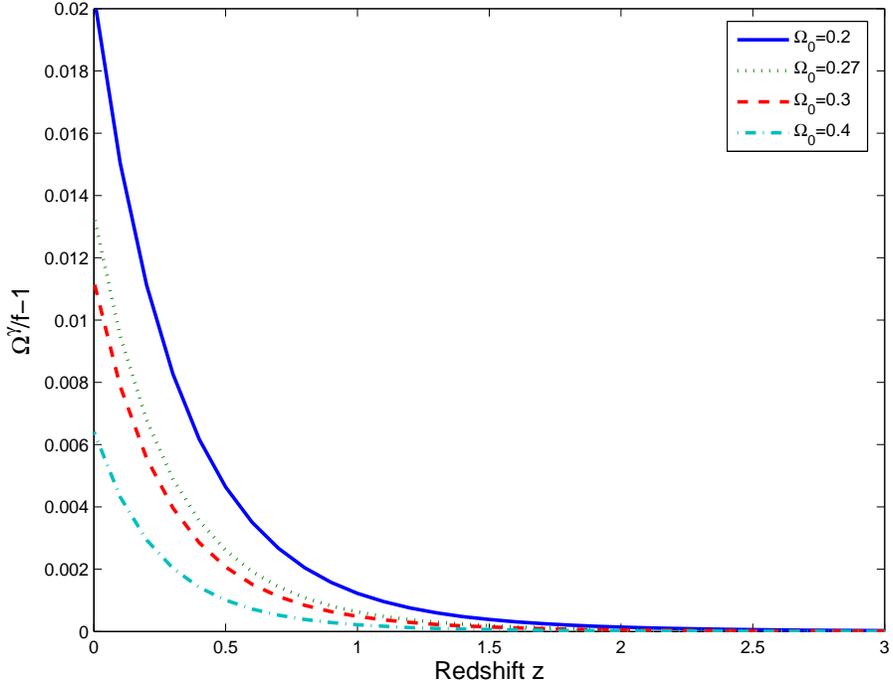}
\caption{The relative difference between the growth factor $f$ and $\Omega^{\gamma_\infty}$ with $\gamma_\infty=6/11$.}
\label{lcdmferr}
\end{figure}

\begin{figure}[htp]
\centering
\includegraphics[width=12cm]{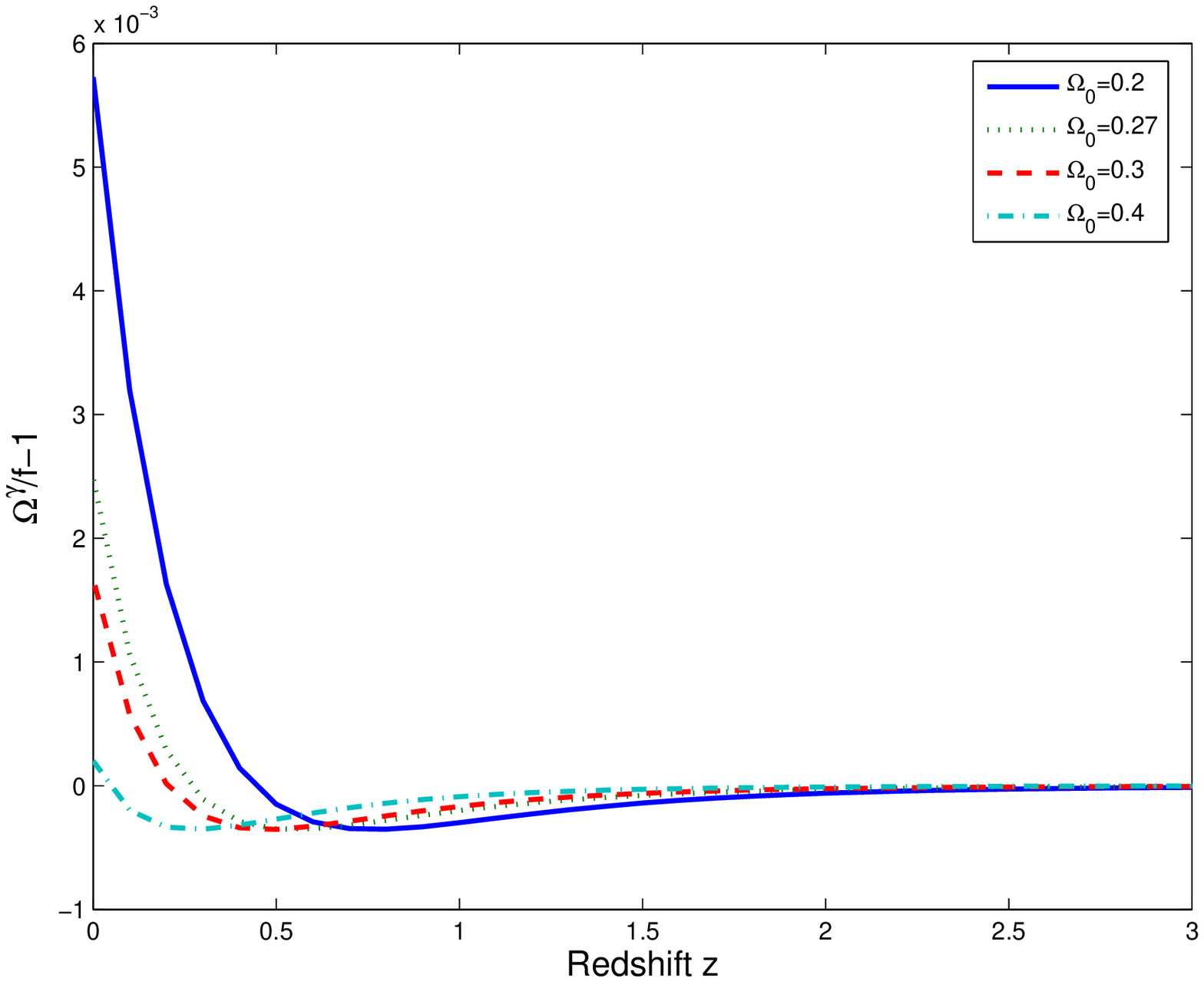}
\caption{The relative difference between the growth factor $f$ and $\Omega^\gamma$ with
$\gamma=\gamma_\infty+\gamma_1 (1-\Omega)$
for the $\Lambda$CDM model.}
\label{lcdmferr1}
\end{figure}

For the $\Lambda$CDM model, $w=-1$, so the growth index becomes
\begin{equation}
\label{lcdmr0}
\gamma_\infty=\frac{3(1-w)}{5-6w}=\frac{6}{11},
\end{equation}
and
\begin{equation}
\label{lcdmr1}
\gamma_1=-\frac{d\gamma}{d\Omega}=\frac{3}{125}\frac{(1-w)(1-3w/2)}{(1-6w/5)^3}=\frac{15}{1331}.
\end{equation}
 For the $\Lambda$CDM model, the matter density is
\begin{equation}
\label{lcdmwm}
\Omega=\frac{\Omega_0 (1+z)^3}{\Omega_0 (1+z)^3+1-\Omega_0}.
\end{equation}
Substituting Eq. (\ref{lcdmwm}) into Eq. (\ref{wcdmfeq1}) and
solving the equation numerically, we compare the numerical result
$f$ with the analytical approximation $\Omega^\gamma$. The results
are shown in Figs. \ref{lcdmferr} and \ref{lcdmferr1}. Since
$\gamma_1$ is very small, $\gamma$ does not change much, and we can
take the approximation $\gamma=\gamma_\infty$. In Fig.
\ref{lcdmferr}, we compare $\Omega^{\gamma_\infty}$ with $f$ for
different values of $\Omega_0$. From Fig. \ref{lcdmferr}, we see
that $\Omega^{\gamma_\infty}$ approximates $f$ very well; the
smaller $\Omega_0$, the larger the error. When the Universe deviates
farther from the matter dominated era, the error becomes larger. For
$\Omega_0=0.2$, the approximation overestimates $f$ by only 2\%, or
$\gamma_\infty$ underestimates $\gamma$. To get a better
approximation, we expand to the first order of $(1-\Omega)$ and use
$\gamma=\gamma_\infty+\gamma_1 (1-\Omega)$. In Fig. \ref{lcdmferr1},
we plot the relative difference between $\Omega^\gamma$ and $f$.
From Fig. \ref{lcdmferr1}, we see that using
$\gamma=\gamma_\infty+\gamma_1 (1-\Omega)$ approximates the growth
factor much better; now the error is only 0.6\% for $\Omega_0=0.2$.

\section{DGP model}

\begin{figure}[htp]
\centering
\includegraphics[width=12cm]{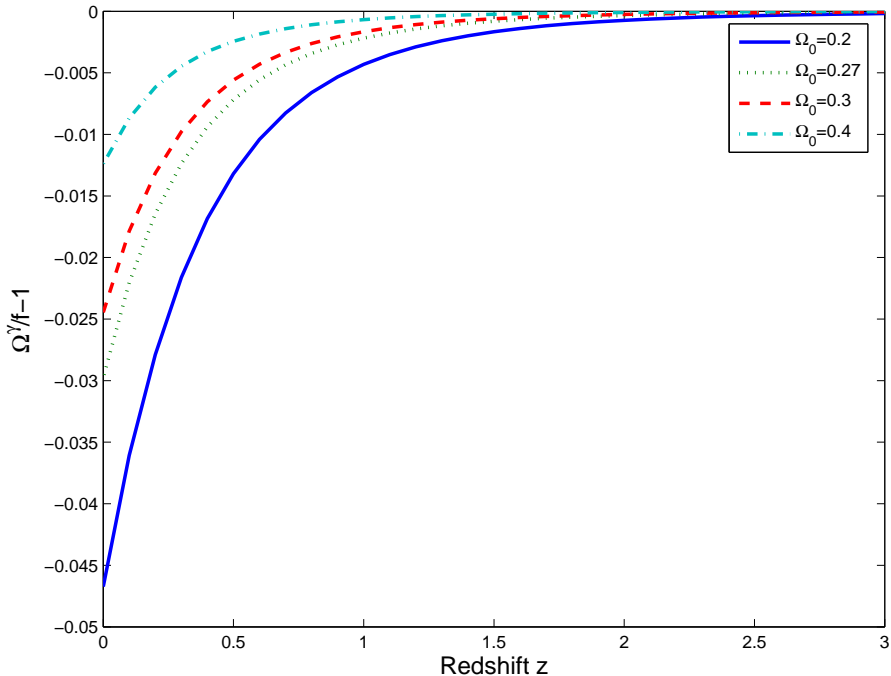}
\caption{The relative difference between the growth factor $f$ and $\Omega^{\gamma_\infty}$
with $\gamma_\infty=11/16$ for the DGP model.}
\label{dgpferr}
\end{figure}

\begin{figure}[htp]
\centering
\includegraphics[width=12cm]{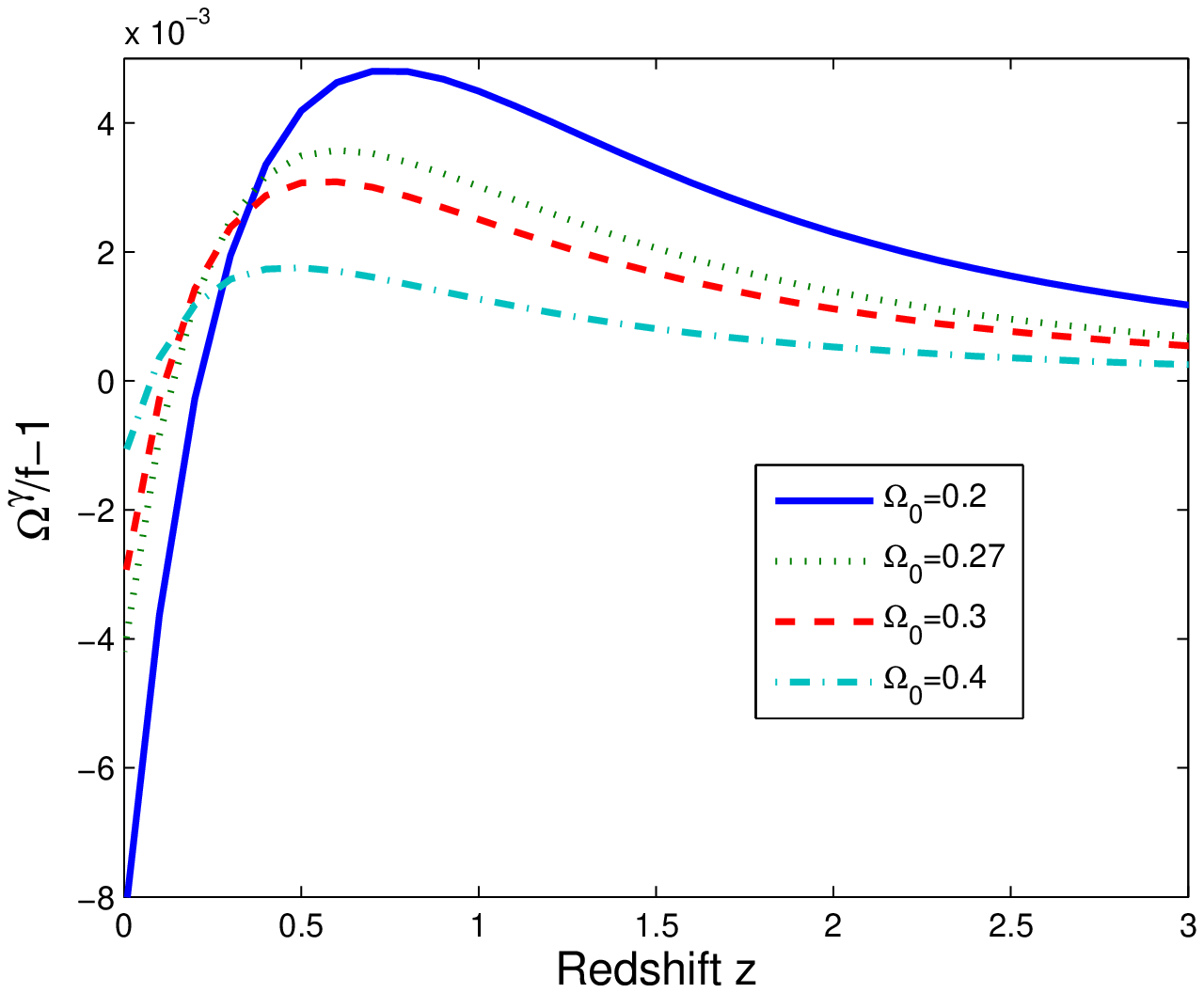}
\caption{The relative difference between the growth factor $f$ and $\Omega^\gamma$
with $\gamma=\gamma_\infty+\gamma_1(1-\Omega)$ for the DGP model.}
\label{dgpferr1}
\end{figure}

For the flat DGP model, we have
\begin{equation}
\label{dgpgeff}
\frac{G_{eff}}{G}=\frac{2(1+2\Omega^2)}{3(1+\Omega^2)}.
\end{equation}
The Friedmann equation tells us that
\begin{equation}
\label{dgphdot}
\frac{\dot H}{H^2}=-\frac{3\Omega}{1+\Omega}.
\end{equation}
The energy conservation equation tells us that
\begin{equation}
\label{dgpwmder}
\Omega'=-\frac{3\Omega(1-\Omega)}{1+\Omega}.
\end{equation}
The matter energy density is given by
\begin{equation}
\label{dgpwm}
\Omega=\frac{\Omega_0(1+z)^3}{[(1-\Omega_0)/2+
\sqrt{\Omega_0 (1+z)^3+(1-\Omega_0)^2/4}\,]^2}.
\end{equation}
Substituting Eqs. (\ref{dgpgeff}), (\ref{dgphdot}) and
(\ref{dgpwmder}) into Eq. (\ref{grwthfeq1}), we get
\begin{equation}
\label{dgpfeq}
-\frac{3\Omega(1-\Omega)}{1+\Omega}\frac{df}{d\Omega}+f^2+\frac{2-\Omega}{1+\Omega}\,f
=\frac{\Omega(1+2\Omega^2)}{1+\Omega^2}.
\end{equation}
Plugging $f=\Omega^\gamma$ into Eq. (\ref{dgpfeq}), we get
\begin{equation}
\label{dgpfeq1}
-\frac{3\Omega(1-\Omega)\ln\Omega}{1+\Omega}\frac{d\gamma}{d\Omega}-\frac{3(1-\Omega)\gamma}{1+\Omega}
+\Omega^\gamma+\frac{2-\Omega}{1+\Omega}-\frac{\Omega^{1-\gamma}(1+2\Omega^2)}{1+\Omega^2}=0.
\end{equation}
Expanding Eq. (\ref{dgpfeq1}) around $\Omega=1$, to the first order
of $(1-\Omega)$, we get
\begin{equation}
\label{dgpr}
\gamma=\frac{11}{16}+\frac{7}{5632}(1-\Omega).
\end{equation}
So $\gamma_\infty=11/16$ and $\gamma_1=7/5632$. The change of
$\gamma$ is very small because $\gamma_1$ is very small; we first
approximate $\gamma$ by $\gamma_\infty$ and the result is shown in
Fig. \ref{dgpferr}. From Fig \ref{dgpferr}, we see that the error
becomes larger when the Universe deviates farther from the matter
domination. When $\Omega_{0}=0.2$, $\Omega^{\gamma_\infty}$
underestimates $f$ by 4.6\%. So the growth factor is overestimated.
If we use the first order approximation, the error becomes larger
because $\gamma_1>0$. Linder and Cahn give the approximation
\cite{linder07}
\begin{equation}
\label{dgpr07}
\gamma\approx \frac{7+5\Omega+7\Omega^2+3\Omega^3}{(1+\Omega^2)(11+5\Omega)}
\approx \frac{11}{16}+\frac{7}{256}(1-\Omega).
\end{equation}
Again, to the first order approximation of $(1-\Omega)$, the error
becomes larger than that with $\gamma_\infty$. In \cite{wei}, the
author found the approximation
\begin{equation}
\label{dgprwei}
\gamma= \frac{11}{16}-\frac{3}{16}(1-\Omega)+\frac{1}{16}(1-\Omega)^2.
\end{equation}
In this approximation, the correction to $\gamma_\infty$ is too big
because $\gamma_1$ is too large even though the sign is correct. To
get better than 1\% approximation, we first assume $\gamma$ is a
constant and solve Eq. (\ref{dgpfeq1}) at $\Omega=\Omega_0$ to get
$\gamma_0$; we then approximate
$\gamma_1=(\gamma_0-\gamma_\infty)/2(1-\Omega_0)$. The values of
$\gamma_0$ and $\gamma_1$ for different $\Omega_0$ are listed in
table \ref{dgpr0r1}. The difference between $\Omega^\gamma$ with
$\gamma=\gamma_\infty+\gamma_1 (1-\Omega)$ and $f$ is shown in Fig.
\ref{dgpferr1}. As promised, the error is under 1\%.

\begin{table}[htp]
\begin{tabular}{|c|c|c|}
\hline
$\Omega_0$ & $\gamma_0$ & $\gamma_1$ \\
\hline
$0.2$ & 0.639 & -0.0305 \\
$0.27$ & 0.648 & -0.0272 \\
$0.3$ & 0.652 & -0.0257 \\
$0.4$ & 0.663 & -0.0205 \\
\hline
\end{tabular}
\caption{The values of $\gamma_0$ and $\gamma_1$ in the DGP model
for different $\Omega_0$.} \label{dgpr0r1}
\end{table}

\section{Observational Constraints}

Now we use the observational data to fit the dark energy model with
constant $w$ and the DGP model. The parameters in the models are
determined by minimizing
$\chi^2=\chi^2_{sn}+\chi^2_a+\chi^2_r+\chi^2_h+\chi^2_f$. For the
SNe data, we use the reduced Union compilation of 307 SNe
\cite{union}. The SNe compilation includes the Supernova Legacy
Survey \cite{astier} and the ESSENCE Survey \cite{riess,essence},
the older observed SNe data, and the extended dataset of distant SNe
observed with the Hubble space telescope. To fit the SNe data, we
define
\begin{equation}
\label{chi}
\chi^2_{sn}=\sum_{i=1}^{307}\frac{[\mu_{obs}(z_i)-\mu(z_i)]^2}{\sigma^2_i},
\end{equation}
where the extinction-corrected distance modulus
$\mu(z)=5\log_{10}[d_L(z)/{\rm Mpc}]+25$, $\sigma_i$ is the total
uncertainty in the SNe data, and the luminosity distance is
\begin{equation}
\label{lum}
d_L(z)=H^{-1}_0(1+z)\int_0^z\frac{dz'}{E(z')}.
\end{equation}
The dimensionless Hubble parameter
$E(z)=H(z)/H_0=[\Omega_0(1+z)^3+(1-\Omega_0)(1+z)^{3(1+w)}]^{1/2}$
for the dark energy model with constant $w$ and
$E(z)=[\Omega_0(1+z)^3+(1-\Omega_0)^2/4]^{1/2}+(1-\Omega_0)/2$ for
the DGP model. The nuisance parameter $H_0$ is marginalized over
using a flat prior.

To use the BAO measurement from the SDSS data, we define
$\chi^2_a=(A-0.469(0.96/0.98)^{-0.35})^2/0.017^2$, where the
distance parameter \cite{sdss6,wmap5}
\begin{equation}
\label{paraa}
A=\frac{\sqrt{\Omega_{0}}}{0.35}\left[\frac{0.35}{E(0.35)}\left(\int^{0.35}_0\frac{dz}{E(z)}\right)^2\right]^{1/3}
=0.469(0.96/0.98)^{-0.35}\pm 0.017.
\end{equation}

To use the shift parameter measured from the WMAP5 data, we define
$\chi^2_r=(\mathcal{R}-1.715)^2/0.021^2$, where the
shift parameter \cite{wmap5}
\begin{equation}
\label{shift}
\mathcal{R}=\sqrt{\Omega_{0}}\int_0^{1089}\frac{dz}{E(z)}=1.715\pm 0.021.
\end{equation}

Simon, Verde, and Jimenez obtained the Hubble parameter $H(z)$ at
nine different redshifts from the differential ages of passively
evolving galaxies \cite{hz1}. Recently, the authors in \cite{hz2}
obtained $H(z=0.24)=83.2\pm 2.1$ and $H(z=0.43)=90.3\pm 2.5$ by
taking the BAO scale as a standard ruler in the radial direction. To
use these 11 $H(z)$ data, we define
\begin{equation}
\label{hzchi}
\chi^2_h=\sum_{i=1}^{11}\frac{[H_{obs}(z_i)-H(z_i)]^2}{\sigma_{hi}^2},
\end{equation}
where $\sigma_{hi}$ is the $1\sigma$ uncertainty in the $H(z)$ data.
We also add the prior $H_0=72\pm 8$ km/s/Mpc given by Freedman {\it
et al.} \cite{freedman}.

For the growth factor data, we define
\begin{equation}
\label{fzchi}
\chi^2_f=\sum_{i=1}^{12}\frac{[f_{obs}(z_i)-\Omega^\gamma(z_i)]^2}{\sigma_{fi}^2},
\end{equation}
where $\sigma_{fi}$ is the $1\sigma$ uncertainty in the $f(z)$ data.
For reference, we compile the available data \cite{porto,ness,guzzo}
in Table \ref{fzdata}. The data are obtained from the measurement of
the redshift distortion parameter $\beta=f/b$, where the bias factor
$b$ measures how closely galaxies trace the mass density field. Note
that some of the measured redshift distortion parameter $\beta$ is
obtained by fitting $\Omega^{0.6}$ with $\Omega$ given by the
$\Lambda$CDM model, and some analyses tried to account for extra
distortions due to the geometric Alcock-Paczynski effect
\cite{guzzo}. With these caveats in mind, it is still worthwhile to
apply the data to fit the models. As discussed in the previous
sections, we can use $\gamma=\gamma_\infty$ within the accuracy of a
few percent. For the $\Lambda$CDM model, we use $\gamma=6/11$ and
$\Omega$ given by Eq. (\ref{lcdmwm}). For the DGP model, we use
$\gamma=11/16$ and $\Omega$ given by Eq. (\ref{dgpwm}).

\begin{table}[htp]
\begin{tabular}{|lcr|}
\hline
\ \ \ \ \ \ \ \ \   $z$\ \ \ \ \ \ \ \ \   & \ \ \ \ \ \ \  $f_{obs}$\ \ \ \ \ \ \  &\ \ References \\
\hline
$0.15$ & $0.49\pm 0.1$ & \cite{colless,guzzo} \\
$0.35$ & $0.7\pm 0.18$ & \cite{tegmark} \\
$0.55$ & $0.75\pm 0.18$ & \cite{ross} \\
$0.77$ & $0.91\pm 0.36$ & \cite{guzzo} \\
$1.4$ & $0.9\pm 0.24$ & \cite{angela} \\
$3.0$ & $1.46\pm 0.29$ & \cite{mcdonald} \\
$2.125-2.72$ & $0.74\pm 0.24$ & \cite{viel1} \\
$2.2-3$ & $0.99\pm 1.16$ & \cite{viel2}\\
$2.4-3.2$ & $1.13\pm 1.07$ & \cite{viel2}\\
$2.6-3.4$ & $1.66\pm 1.35$ & \cite{viel2} \\
$2.8-3.6$ & $1.43\pm 1.34$ & \cite{viel2} \\
$3-3.8$ & $1.3\pm 1.5$ & \cite{viel2}\\
\hline
\end{tabular}
\caption{The summary of the observational data on the growth factor
$f$.}
\label{fzdata}
\end{table}

\begin{figure}[htp]
\centering
\includegraphics[width=12cm]{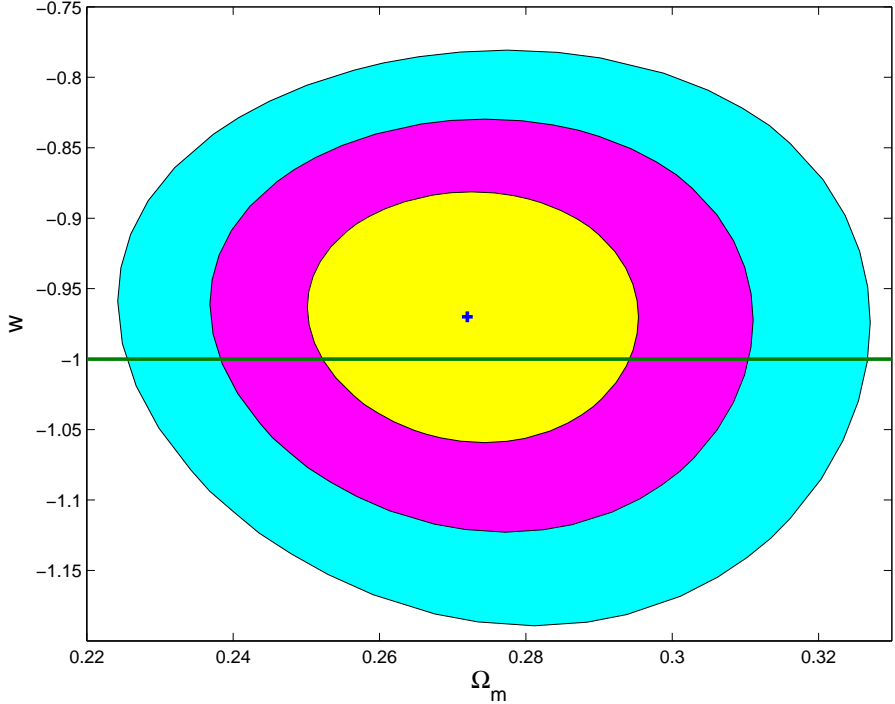}
\caption{The $1\sigma$, $2\sigma$, and $3\sigma$ contours of $\Omega$ and $w$ by fitting
the dark energy model with constant $w$ to the combined data. The point with $+$
denotes the best fit value.}
\label{omwcont}
\end{figure}

By fitting the dark energy model with constant $w$ to the combined
data, we get $\chi^2=325.17$, $\Omega_0=0.272^{+0.023}_{-0.022}$,
and $w=-0.97\pm 0.09$. The $1\sigma$, $2\sigma$, and $3\sigma$
contours of $\Omega_0$ and $w$ are shown in Fig. \ref{omwcont}. From
Fig. \ref{omwcont}, we see that the $\Lambda$CDM model is consistent
with the current observation. By fitting the $\Lambda$CDM model to
the combined data, we get $\chi^2=325.48$ and $\Omega_0=0.273\pm
0.015$. By fitting the DGP model to the combined data, we get
$\chi^2=350.12$ and $\Omega_0=0.278\pm 0.015$.

\begin{figure}[htp]
\centering
\includegraphics[width=12cm]{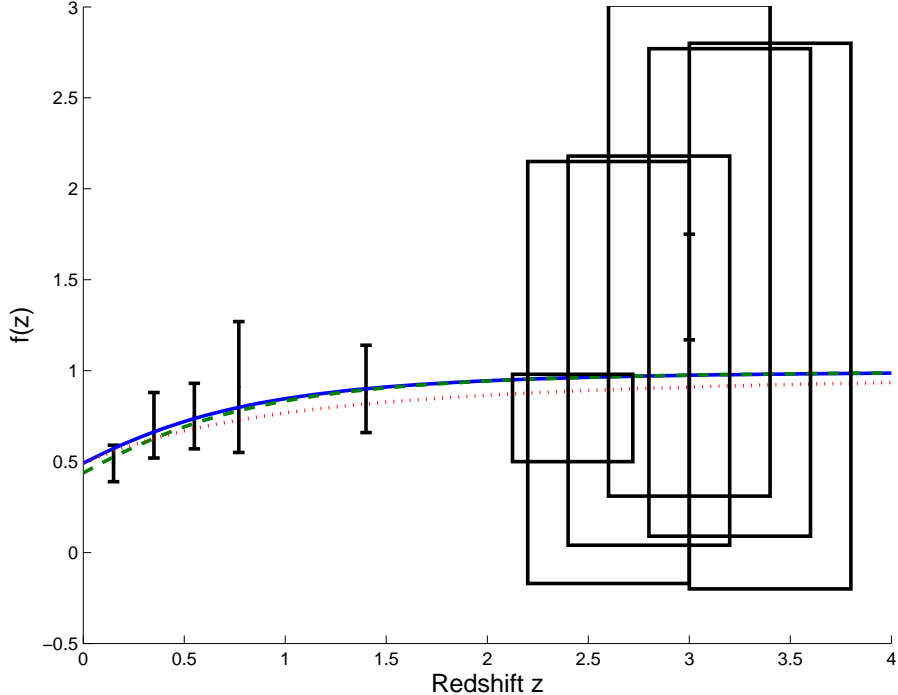}
\caption{The growth factor data and the best fit curves. The solid line is for
the dark energy model with $\Omega_0=0.272$ and $w=-0.97$. The dashed line is
for the $\Lambda$CDM model with $\Omega_0=0.273$ and $\gamma=0.64$. The dotted
line is for the DGP model with $\Omega_0=0.278$ and $\gamma=0.55$.}
\label{fzplot}
\end{figure}

If we fit $\Omega^\gamma$ to the growth factor data $f(z)$ alone, we
can get a constraint on the growth index $\gamma$. For the
$\Lambda$CDM model with the best fit value $\Omega_0=0.273$, we find
that $\chi^2=4.55$ and $\gamma_\Lambda=0.64^{+0.17}_{-0.15}$. The
theoretical value $\gamma_\infty=6/11=0.55$ is consistent with the
observation at the $1\sigma$ level. For the DGP model with the best
fit value $\Omega_0=0.278$, we find $\chi^2=5.68$ and
$\gamma_{DGP}=0.55^{+0.14}_{-0.13}$ which is barely consistent with
the theoretical value $\gamma_\infty=11/16=0.6875$ at the $1\sigma$
level. The best fit curves $\Omega^\gamma$ and the observational
data are shown in Fig. \ref{fzplot}.

\section{Discussions}

The simple analytical formulas $\Omega^\gamma$ can be used to
approximate the growth rate $f$. The value of $\gamma$ provides
useful information about the dark energy model and the modification
of gravity. For the dark energy model within general relativity,
$\gamma\approx 0.55$. For the DGP model, $\gamma=0.6875$. If the
accuracy of the growth factor data is in the range of a few percent,
we can use a constant $\gamma_\infty$ to approximate $f$. For
example, $\gamma_\infty=3(1-w)/(5-6w)$ for constant $w$ \cite{wang},
$\gamma_\infty=0.55+0.05[1+w(z=1)]$ for dynamical models
\cite{linder05}, and $\gamma_\infty=11/16$ for the DGP model. The
value of $\gamma_\infty$ is obtained by approximating the
differential equation of $f$ around $\Omega=1$. This approximation
is reasonably good at high redshift ($z\gtrsim 1$) when
$\Omega\approx 1$. However, at lower redshift the dark energy or the
effect of modified gravity starts to dominate and $\Omega$ deviates
from 1; we expect the approximation to break down. Therefore, to get
a better than 1\% fit, the $\Omega$ dependence of $\gamma$ needs to
be considered. We show that $\gamma=\gamma_\infty+\gamma_1
(1-\Omega)$ can approximate $f$ to better than 1\%. The value of
$\gamma_1$ is very small compared with $\gamma_\infty$ and usually
depends on the value of $\Omega_0$. For the $\Lambda$CDM model,
$\gamma_1=15/1331$. For the DGP model, we give a prescription of how
to find $\gamma_1$; the values of $\gamma_1$ are listed in Table
\ref{dgpr0r1} for some values of $\Omega_0$.

To distinguish different dark energy models and modified gravity, we
use the observational data to fit the models. Fitting the combined
SNe, SDSS, WMAP5, $H(z)$, and $f(z)$ data to the dark energy model
with constant $w$, we find that $\chi^2=325.17$,
$\Omega_0=0.272^{+0.023}_{-0.022}$, and $w=-0.97\pm 0.09$. For the
$\Lambda$CDM model, we find that $\chi^2=325.48$ and
$\Omega_0=0.273\pm 0.015$. For the DGP model, we find that
$\chi^2=350.12$ and $\Omega_0=0.278\pm 0.015$. The results suggest
that the data strongly disfavor the DGP model. If we fit the SNe
data alone to the DGP model, we get $\chi^2=313.27$ and
$\Omega_0=0.186$. This result is consistent with the results in
\cite{song,lue1,song05,koyama,roy06,gong,zhu}. If the same data were
fitted to the $\Lambda$CDM model, we would get $\chi^2=312.18$. If
we fit the $\Lambda$CDM model to the combined SNe and $H(z)$ data,
we get $\chi^2=319.45$. For the DGP model, we get $\chi^2=321.31$
and $\Omega_0=0.194$. These results show that both the $\Lambda$CDM
model and the DGP model give almost the same expansion rate. If we
fit the $\Lambda$CDM model to the combined SNe, $H(z)$, and $f(z)$
data, we get $\chi^2=324.53$. For the DGP model, we get
$\chi^2=331.53$ and $\Omega_0=0.207$. With the addition of $f(z)$
data to the expansion data, the DGP model is readily distinguishable
from the $\Lambda$CDM model. As mentioned above, the best fit value
of $\Omega_0$ in the DGP model tends to be lower, i.e.,
$\Omega_0\approx 0.2$, when we fit the model to the SNe, $H(z)$, and
$f(z)$ data. On the other hand, we get $\chi^2=1.4$ and
$\Omega_0=0.33$ if we fit the model to the distance parameter $A$
and the shift parameter $\mathcal{R}$. If the two sets of data are
combined together, $\Omega_0$ takes the value in the middle and we
get a large value of $\chi^2$. When we fit the models to the
combined SNe, the distance parameter $A$, the shift parameter
$\mathcal{R}$, and $H(z)$ data, we get $\Omega_0=0.274$ and
$\chi^2=320.6$ for the $\Lambda$CDM model and $\Omega_0=0.275$ and
$\chi^2=343.5$ for the DGP model. This result is consistent with the
analysis by Song {\it et al} \cite{song}. In \cite{song}, they found
that the flat DGP model is excluded at about $3\sigma$ by the
combined SNe, 3 yr WMAP, and Hubble constant data.

The observational data of $f(z)$ can be used to provide information
on the growth index $\gamma$ and the modified gravity. As discussed
above, $\gamma$ is almost a constant; we can use $\Omega^\gamma$
with a constant $\gamma$ to approximate $f(z)$. For the $\Lambda$CDM
model, we find that $\gamma_\Lambda=0.64^{+0.17}_{-0.15}$, which is
consistent with the theoretical value 0.55. This result is also
consistent with that in \cite{ness,porto}. For the DGP model, we
find that $\gamma_{DGP}=0.55^{+0.14}_{-0.13}$. The theoretical value
$\gamma=0.6875$ lies on the upper limit of the $1\sigma$ error. From
Fig. \ref{fzplot}, we see that the growth rates for the DGP model
and the $\Lambda$CDM model are distinguishable even with the best
fitting values of $\gamma$. If the theoretical values of $\gamma$
are used, the difference will be larger. The approximation of
$\Omega^\gamma$ to $f$ is good for theories with $f\le 1$ and
$\Omega\le 1$. If the observational data for $f(z)$ are larger than
1 at high redshift ($z\gtrsim 1$), then the approximation
$\Omega^\gamma$ is broken and the effect of modified gravity is
explicit. For the Brans-Dicke theory, during the matter domination,
$\Omega=(3\omega+4)(2\omega+3)/6(1+\omega)^2\ge 1$ and
$f=(2+\omega)/(1+\omega)\ge 1$ \cite{porto}; here $\omega$ is the
Brans-Dicke constant. In this case, modification of the
approximation is needed. This will be discussed in future work. In
conclusion, more precise future data on $f(z)$ along with the SNe
data will differentiate dark energy models from modified gravity.

\begin{acknowledgments}
This research was supported in part by the Project of Knowledge
Innovation Program (PKIP) of the Chinese Academy of Sciences and by
NNSFC under Grant No. 10605042. The author thanks the hospitality of
Kavli Institute for Theoretical Physics China and the Abdus Salam
International Center for Theoretical Physics where part of the work
was done.
\end{acknowledgments}

\end{document}